\newcommand{\za}{\alpha}

\newcommand{\zs}{\sigma}

\newcommand{\ze}{\epsilon}

\newcommand{\zm}{\mu}

\newcommand{\zt}{\tau}

\newcommand{\zF}{\Phi}
\newcommand{\zG}{\Gamma}

\newcommand{\zR}{I\hskip-3.4pt R}

\newcommand{\zW}{\Omega}

\documentstyle[11pt]{article}

\newlength{\fullboxwidth}
\begin{document}

\title{NOTE ON PHASE SPACE CONTRACTION AND ENTROPY PRODUCTION IN
THERMOSTATTED HAMILTONIAN SYSTEMS}
\author{
E.G.D. Cohen \\
Rockefeller University, New York, New York 10021 - U.S.A.\\
\\
L. Rondoni \\ Dipartimento di Matematica, Politecnico di Torino\\
Corso Duca degli Abruzzi 24, I-10129 Torino, Italy \\ \\
e-mail:  rondoni @ polito.it \\ \\ \\
Physics abstracts numbers: 05.45.+b, 05.60.+w, 05.70.Ln}

\maketitle



\noindent
{\Large \bf Abstract}
\vskip 10pt \noindent
The phase space contraction and the entropy production rates of
Hamiltonian systems in an external field, thermostatted to obtain
a stationary state are considered. While for stationary states
with a constant kinetic energy the two rates are formally
equal for all numbers of particles $N$, for stationary states
with constant total (kinetic and potential) energy this only
obtains for large $N$. However, in both cases a large number 
of particles is required to obtain equality with the entropy 
production rate of Irreversible Thermodynamics. Consequences
of this for the positivity of the transport coefficients 
and for the Onsager relations are discussed. Numerical results 
are presented for the special case of the Lorentz gas.

\bigskip

\section{Introduction}
In the last decade or so, important new connections have been made
between the mathematical theory of dynamical systems and the 
statistical mechanics of systems in nonequilibrium stationary
states.
In particular, the average phase space contraction rate
of a dissipative dynamical system was equated to the physical entropy
production rate as given by Irreversible Thermodynamics. This has also 
led to apparent contradictions between the mathematical theorems
established on the basis of dynamical systems theory and the lore of
statistical mechanics about irreversibility. 
The purpose of this paper is to reconcile
these apparently contrasting results.

For instance, the proof that the entropy production rate\footnote{We should 
rather say {\it generalized} entropy production, because the stationary
states we consider may be quite far from equilibrium. We will not do this,
in order to simplify the language, so {\it generalized} should 
be understood where appropriate.}
in a dynamical system in a nonequilibrium stationary state is 
necessarily positive \cite{RU} (see also Refs.\cite{RMlarge,RC}),
would imply that certain transport 
coefficients are also positive, in accordance with the Second
Law of Thermodynamics. In other words, the Law could be directly verified
for such a dynamical system. Because the proof in Ref.\cite{RU}
only needs the system to be strongly chaotic and dissipative, with no 
vanishing Lyapunov exponents, it 
appears that the number of particles $N$ plays no role in the 
derivation of the Second Law. This contradicts one of the basic tenets 
of statistical mechanics, which attributes the Second Law of
Thermodynamics to the very large number of particles present
in macroscopic systems.\footnote{A similar argument was used in the
context of Hamiltonian systems by J.L. Lebowitz in {\it Boltzmann's 
entropy and time's arrow}, Physics Today, September 1993, p.38.
The paper ended with this statement:
{\it This is an essential difference (unfortunately frequently overlooked
or misunderstood) between the irreversible and the chaotic behavior of
Hamiltonian systems. The latter, which can be observed already in systems
consisting of only a few particles, will not have unidirectional time
behavior in any particular realization. Thus if we had only a few
hard spheres in a box, we would get plenty of chaotic dynamics and very
good ergodic behavior, but we could not tell the time order of any
sequence of snapshots.}}

Similarly, Ref.\cite{GR} gives a proof the
validity of Onsager reciprocal relations (OR), and
explicitly states that the proof is valid for any number
of particles. This is at least in appearance at variance with
Onsager's derivation of the OR, where he considers macroscopic
systems containing a sufficiently large number of particles, that they
can be subdivided into subsystems, still containing a very large
number of particles (see Refs.\cite{RC,GG1,GG2} for other recent
derivations of the OR).

In this paper, we will attempt to reconcile these apparent paradoxes,
along, in fact, similar lines as Boltzmann tried to reconcile
Zermelo's recurrence paradox, based on Poincar\'e's recurrence theorem
for dynamical systems, with his own results for the approach to
equilibrium, based on the H-theorem. The main point here is to
distinguish between mathematical correctness and physical
relevance. There is no question that
the results mentioned above, and others as well, which are solely based 
on the strong chaoticity of the relevant dynamical systems, are
correct. However, in this paper we argue
that they acquire physical significance only
when they pertain to ``macroscopic'' systems,
i.e. to dynamical systems with many degrees of freedom, representing
the dynamics of a large number of particles. 

To make this idea more precise, we focus on a special class of 
systems consisting
of $N$ particles in a $d$ dimensional space, whose coordinates
and momenta are denoted respectively by ${\bf q} = \{{\bf q}_i\}_{i=1}^N$, and 
${\bf p} = \{{\bf p}_i\} _{i=1}^N$. The relevant phase space is taken to be
$\zW \subset \zR^{2dN}$ and a point $\zG \in \zW$ represents
a possible phase for the system. Where appropriate we use 
$({\bf q},{\bf p})$ to represent a point in $\zW$.
In general, the phase of the system
changes in time defining an evolution in $\zW$, which we represent by
the one paramter group of transformations $S^t : \zW \rightarrow \zW$, 
such that $S^t \zG$ is the phase of the system at the time $t \in \zR$, 
if the initial phase is $\zG$ ($S^0 \zG = \zG$). The special class of 
dynamical systems of this form that we consider are the isokinetic (IK)
as well as the isoenergetic (IE) thermostatted Hamiltonian systems,
to be described in Section 2.
A dynamical system of this
kind will be called ``macroscopic'' if $N$ is sufficiently large
that its properties can also be observed in physical systems, 
i.e. in real experiments.\footnote{One may be satisfied with
the possibility of observing a given property in a 
numerical simulation. In this case, the property must become evident
within a limited number of steps, also to prevent
the numerical noise from overwhelming the measured quantity. 
For instance, the CPU time for a simulation of $N$ non interacting
particles for a time $t$ may be the same as that for a simulation of 
$2 N$ particles for a time $t/2$, but the numerical errors  
may be smaller in the second case. Hence higher $N$
should still be preferred. More importantly, our point concerns  
physical times not the CPU times of computer experiments, even if 
these can be used in place of real experiments.}

The motivation for saying that the properties of such
systems have a physical meaning only for large $N$ 
is at least two-fold. In the first place,
it should be borne in mind that physical (measurable)
quantities are defined by time averages, such that the measured
quantity corresponding to a phase variable $A$ is given by
\begin{equation}
\bar{A}(\zG) = \lim_{T\rightarrow\infty} A_T(\zG)~, \quad A_T(\zG) \equiv
\frac{1}{T} \int_0^T
A(S^\tau \zG) d \zt ~.
\label{def1}
\end{equation}
If one considers a class of sufficiently regular functions $A$,
the notion of ergodicity can be introduced, which implies the 
existence of a probability measure $\zm$ on $\zW$ such that
$\bar{A}(\zG) = \langle A \rangle$ for almost all $\zG \in \zW$, where
\begin{equation}
\langle A \rangle = \int_\zW A(\zG) d \mu(\zG) ~.
\label{def2}
\end{equation}
The measure $\mu$ then represents the stationary state of the system, is
$S^t$ invariant, and is called natural (or physical) measure. 
The existence of such a measure is usually only assumed, because 
the cases of interest in statistical mechanics for which a proof has 
been given are very few (see Ref.\cite{SA} for a recent review on the 
subject). Thus,
we also assume that the systems under consideration are ergodic.
Secondly, however, the presence of a limit in $T$
in Eq.(\ref{def1}) raises the question of which are the typical
time scales of the system. In other words, one should be able to identify 
the time $T$ that is needed in order for the difference between
$A_T$ and $\langle A \rangle$ to become a negligible quantity. This value 
of $T$ is then the relaxation time of the system and to make 
physical sense this time should roughly be of the order of the physical 
times, which are the collision or interaction times, of the system. 
We say that a necessary condition for a system to be considered macroscopic 
(or physically relevant) is that its relaxation time remains within 
physical times. This view implies, in particular, that the equality of 
the average phase space contraction and entropy production rates used 
in the papers mentioned above may be physically
relevant only for macroscopic systems, because only then 
will the relaxation times remain within physical times.

In Section 4. the paradigmatic case of a Lorentz gas is treated in detail, 
to give an example in support of the view that only a system consisting 
of many particles does, under general conditions, approach the infinite
time limit sufficiently fast that its properties 
can be observed in experiments or, at least, in numerical simulations. 
In Section 2 and Section 3, we show that there is another reason 
for requiring that $N$ be large. 
Indeed, the entropy production and the phase space
contraction rates of the IE thermostatted Hamiltonian systems 
differ by terms of order $O(1/N)$. 
Thus, another necessary condition for a system to be 
considered physically relevant is that such terms are sufficiently 
small. In Section 5. a discussion of 
the results is given. In particular, in that section we point out
the connection of our results with the problem of equivalence of
ensembles in nonequilibrium statistical mechanics.

\section{Thermostatted Hamiltonian systems} 
The equations of motion for a thermostatted Hamiltonian system
in $d$-dimensional space are:
\begin{eqnarray}
\dot{\bf q}_i &=& \frac{{\bf p}_i}{m} ~,  \nonumber \\
\dot{\bf p}_i &=& -\frac{\partial}{\partial {\bf q}_i} \left[
\zF^{\mbox{int}}({\bf q}) + \zF^{\mbox{ext}}({\bf q}) \right] - 
\za(\zG) {\bf p}_i \label{eqsmot} \\ 
&=& {\bf F}_i^{\mbox{int}}({\bf q}) + {\bf F}_i^{\mbox{ext}}({\bf q}) - 
\za(\zG) {\bf p}_i ~, \quad i = 1, ..., N ~. \nonumber
\end{eqnarray}
where we assume that the $N$ particles have the same mass $m$,
$\zF^{\mbox{int}}({\bf q})$ is the potential energy of the
interactions among the particles and $\zF^{\mbox{ext}}({\bf q})$
is the potential of the external fields of force doing work on 
the system. The gradients of such potentials give
the forces acting on the particles, which we denote by
${\bf F}_i^{\mbox{int}}({\bf q})$ and 
${\bf F}_i^{\mbox{ext}}({\bf q})$.
The ``friction'' term, $- \za(\zG) {\bf p}_i$, represents a
coupling with a thermostat, which allows the system to reach
a stationary state, despite the presence of an
external field of force.

The stationary state is usually obtained in one of two ways: either
one imposes an isokinetic (IK) constraint, so that the total kinetic 
energy of the system $K = \sum_{i=1}^N {\bf p}_i^2/2 m$ remains 
constant, or an isoenergetic (IE)
constraint is imposed so that the total internal energy of the system
$E = K + \zF^{\mbox{int}}$ remains constant.
The two constraints result, in general, into two different 
stationary states. There are various ways of imposing that the equations 
of motion of our systems obey one or the other (non-holonomic) constraint. 
Here, we follow Gauss' principle of least constraint 
\cite{GA,EM}, obtaining the following expressions
for the term $\za$ in Eqs.(\ref{eqsmot}):
\begin{equation}
\za_{_{IK}}(\zG) = 
\frac{\sum_{i=1}^N {\bf p}_i \cdot \left[
{\bf F}_i^{\mbox{int}}({\bf q}) + {\bf F}_i^{\mbox{ext}}({\bf q})\right]}
{\sum_{i=1}^N {\bf p}_i^2}
\label{alfaIK}
\end{equation}
for the IK constraint and 
\begin{equation}
\za_{_{IE}}(\zG) = 
\frac{\sum_{i=1}^N {\bf p}_i \cdot {\bf F}_i^{\mbox{ext}}({\bf q})}{ 
\sum_{i=1}^N {\bf p}_i^2}
\label{alfaIE}
\end{equation}
for the IE constraint, respectively. 
Substituting either Eq.(\ref{alfaIK}) or
Eq.(\ref{alfaIE}) into Eqs.(\ref{eqsmot}) assures that the stationary
states will be characterized by a given kinetic or total energy as 
appropriate. Note that the presence of the constraint reduces
the number of degrees of freedom from $2dN$, for
the non-thermostatted Hamiltonian system, to
$2dN-1$. We now compute for the
systems given by Eqs.(\ref{eqsmot},\ref{alfaIK},\ref{alfaIE}) the
phase space contraction rate defined by:
\begin{equation}
\kappa_{_{IK}}(\zG) \equiv - \frac{d}{d \zG} \cdot \dot{\zG} =
- \sum_{i=1}^N \left[ \frac{d}{d {\bf q}_i} \cdot \dot{\bf q}_i +
\frac{d}{d {\bf p}_i} \cdot \dot{\bf p}_i \right] =
(dN-1) \za_{_{IK}}(\zG)
\label{pscIK}
\end{equation}
for the IK case, and
\begin{equation}
\kappa_{_{IE}}(\zG) \equiv - \frac{d}{d \zG} \cdot \dot{\zG} =
- \sum_{i=1}^N \left[ \frac{d}{d {\bf q}_i} \cdot \dot{\bf q}_i +
\frac{d}{d {\bf p}_i} \cdot \dot{\bf p}_i \right] =
(dN-1) \za_{_{IE}}(\zG)
\label{pscIE}
\end{equation}
for the IE case respectively. In these definitions, we have introduced a 
minus sign anticipating the positivity of $\za$ (on average) 
in the stationary state.

In order to compare the phase space contraction rates
with the entropy production rates of Irreversible Thermodynamics,
we first have to average the phase space contraction rates over
the stationary state, so that they can be 
compared with macroscopic quantities. This leads to 
\begin{equation}
\left\langle \kappa_{_{IK}} \right\rangle_{ss} =
(dN-1) \left[ 
\left\langle \frac{
\sum_{i=1}^N \frac{{\bf p}_i}{m} \cdot {\bf F}_i^{\mbox{int}}
}{
\sum_{i=1}^N \frac{{\bf p}_i^2}{m}
} \right\rangle_{ss} +
\left\langle \frac{
\sum_{i=1}^N \frac{{\bf p}_i}{m} \cdot {\bf F}_i^{\mbox{ext}}
}{
\sum_{i=1}^N \frac{{\bf p}_i^2}{m}
} \right\rangle_{ss}
\right]
\label{KIK}
\end{equation}
where we have divided numerator and denominator by the particles'
mass $m$.

Noting that in the IK case one can define the stationary 
state temperature $T_{ss}$ by:
\begin{equation}
\sum_{i=1}^N \frac{{\bf p}_i^2}{m} = \left\langle 
\sum_{i=1}^N \frac{{\bf p}_i^2}{m} \right\rangle_{ss} =
(dN-1) k_{_B} T_{ss}
\label{TIKss}
\end{equation}
and that
\begin{equation}
\sum_{i=1}^N \frac{{\bf p}_i}{m} \cdot {\bf F}_i^{\mbox{int}}({\bf q}) =
- \frac{d}{d t} \zF^{\mbox{int}}({\bf q})
\label{phider}
\end{equation}
so that 
\begin{equation}
- \left\langle \frac{d}{d t} \zF^{\mbox{int}} \right\rangle_{ss} =
- \frac{d}{d t} \left\langle \zF^{\mbox{int}} \right\rangle_{ss} = 0
\end{equation}
we obtain:
\begin{equation}
\hat{\kappa}_{_{IK}} \equiv \frac{\langle \kappa_{_{IK}} \rangle_{ss}}{V} = 
\frac{\left\langle \sum_{i=1}^N \frac{{\bf p}_i}{m V} \cdot 
{\bf F}_i^{\mbox{ext}} \right\rangle_{ss}}{
k_{_B} T_{ss}} ~.
\label{KIKhat}
\end{equation}
Here, we have introduced $\hat{\kappa}_{_{IK}}$,
by dividing $\langle \kappa_{_{IK}} \rangle_{ss}$
by the volume $V$ of the system, in order to compare later dynamical 
averages with macroscopic thermodynamic quantities. 
Assuming for simplicity that ${\bf F}_i^{\mbox{ext}}({\bf q})$ is 
independent of $i$ and ${\bf q}$: i.e.
${\bf F}_i^{\mbox{ext}}({\bf q})$ is the constant ${\bf F}^{\mbox{ext}}$, 
and noting that 
\begin{equation}
\frac{1}{V} \left\langle \sum_{i=1}^N \frac{{\bf p}_i}{m} \right\rangle_{ss}
\label{currdef}
\end{equation}
is just the particle current density ${\bf J}$, one obtains from 
(\ref{KIKhat}) 
\begin{equation}
\hat{\kappa}_{_{IK}} = \frac{{\bf J} \cdot {\bf F}^{\mbox{ext}}}{
k_{_B} T_{ss}} ~.
\label{KIKhapri}
\end{equation}

Similarly, from Eqs.(\ref{alfaIE},\ref{pscIE}), one has for the IE case:
\begin{equation}
\hat{\kappa}_{_{IE}} \equiv \frac{\langle \kappa_{_{IE}} \rangle_{ss}}{V} = 
(dN-1) \left\langle \frac{
\sum_{i=1}^N \frac{{\bf p}_i}{m V} \cdot 
{\bf F}^{\mbox{ext}}}{
\sum_{i=1}^N \frac{{\bf p}_i^2}{m}} \right\rangle_{ss}
\label{KIEhat}
\end{equation}
However, now only for large $N$ could the average of the ratio on
the right hand side of (\ref{KIEhat}) be replaced by the ratio
of the averages, the latter being typical for macroscopic relations.
Setting then in addition
\begin{equation}
\left\langle \sum_{i=1}^N \frac{{\bf p}_i^2}{m} \right\rangle_{ss} =
d N k_{_B} T_{ss}
\label{tempdef}
\end{equation}
and neglecting terms of order $1/N$ in the ratio $(dN-1)/dN$ as
well, one obtains 
\begin{equation}
\hat{\kappa}_{_{IE}} = \frac{{\bf J} \cdot {\bf F}^{\mbox{ext}}}{
k_{_B} T_{ss}} ~.
\label{KIEhapri}
\end{equation}
The identity between the entropy production and phase space contraction
rates per unit volume requires now that $N$ be large.

\section{Irreversible Thermodynamics}
In order to compare the expressions for $\hat{\kappa}_{_{IK}}$ and
$\hat{\kappa}_{_{IE}}$ with the corresponding entropy density production
rate used in Irreversible Thermodynamics, we should first
realize that in the latter case the system is supposed to be in first
approximation in a (local) thermodynamic equilibrium state, where all
extensive properties are proportional to $N$ and depend further
only on the temperature and the number density $n = N/V$. 
The entropy production rate per unit volume of Irreversible Thermodynamics
is then that for
the (adiabatic) Hamiltonian system defined by Eq.(\ref{eqsmot})
with $\za(\zG)=0$, and given by 
\begin{equation}
\zs = \frac{{\bf J} \cdot {\bf X}}{
k_{_B} T_{ss}} ~.
\label{ITentr}
\end{equation}
Here, {\bf J} is the macroscopic current density induced in the
system by the external force ${\bf F}^{\mbox{ext}}$, $T_{ss}$
is the temperature of the stationary state and {\bf X} is the
thermodynamic force conjugate to the current density {\bf J}.
One would like that the same expression holds both for the IK and IE
cases, i.e. that
\begin{equation}
\zs_{_{IK}} = \zs_{_{IE}} = \zs ~
\label{equent}
\end{equation}
holds. We consider the IK case first.

To compare (\ref{ITentr}) with quantities derivable from
a Hamiltonian dynamical system, we assume that the thermodynamic
force is like an, e.g. electric, field {\bf E}, that can be 
derived from an external potential. Identifying now in
(\ref{KIKhat}) 
\begin{equation}
\left\langle\sum_{i=1}^N \frac{{\bf p}_i}{mV} \cdot {\bf F}^{\mbox{ext}}
\right\rangle_{ss} = {\bf J} \cdot {\bf X}
\label{identif}
\end{equation}
where both sides of the equation give the work done on the
system in the stationary state
by the external force ${\bf F}^{\mbox{ext}} = {\bf X}$
per unit volume and unit time, we see that (\ref{KIKhat},\ref{KIKhapri})
and (\ref{ITentr}) formally agree.

With a similar identification for (\ref{KIEhapri}) and (\ref{ITentr}),
one also has a formal agreement between the two equations for 
the IE case, albeit that the large number of particles in the
system had to be used to obtain this agreement.

\section{The Lorentz gas $-$ Color diffusion}
In this section we use a simple model $-$the thermostatted
Lorentz gas$-$ to show that the formal
agreement obtained between dynamical phase space contraction
and entropy production rates of thermostatted Hamiltonian systems in
nonequilibrium stationary states only has {\em physical} significance
if the number of particles $N$ of these systems is large. In fact, a new
large $N$ requirement comes in addition to that of the previous section
and holds both for IK and IE systems. That is, that the relaxation
times to obtain the observed macroscopic values of the physical
properties of a system might be
unphysically long for systems consisting of a small number of
particles. As an illustration,
we consider numerical results from the equilibrium Lorentz gas with
periodic boundary conditions. The 
relevance of these results for nonequilibrium systems will be
discussed at the end of this section and the beginning of next section
(point 1).

The thermostatted Lorentz gas consists of one point 
particle of mass $m$, which moves through an 
array of fixed scatterers in the presence of a constant
external field {\bf E} driving the particle. We do not
require the scatterers to be hard; they might be soft as in the model 
studied in Ref.\cite{BEC}. For
simplicity, we assume that the particle has a (color) charge
$c$ on which the field {\bf E} acts. The equations of
motion for the moving particle are then:
\begin{eqnarray}
\dot{\bf q} &=& \frac{\bf p}{m }~,  \nonumber \\
\dot{\bf p} &=& 
{\bf F}^{\mbox{int}}({\bf q}) + c {\bf E} - 
\za(\zG) {\bf p} ~, 
\label{eqsLG}
\end{eqnarray}
Here, $-\za(\zG){\bf p}$ represents again a thermostat coupling
needed to attain a stationary state for the moving particle, by
balancing the continuous acceleration of the particle due to the
external field {\bf E} with an effective friction term.

If the scatterers are hard disks or hard spheres, as 
usually considered, there is no internal potential energy
in the system and ${\bf F}^{\mbox{int}}$ is a purely impulsive
force, causing specular reflection of the particle from a
scatterer. In that case, the IK constraint is the same as the
IE constraint, and one has:
\begin{equation}
\za_{_{IK}}(\zG) = \za_{_{IE}}(\zG) = 
\frac{{\bf p} \cdot c {\bf E}}{{\bf p}^2} ~.
\label{alfaLG}
\end{equation}
In general we have then:
\begin{equation}
\left\langle \kappa_{_{IK}} \right\rangle_{ss} =
\frac{\left\langle {\bf p} \cdot \left( {\bf F}^{\mbox{int}} 
+ c {\bf E} \right) / m \right\rangle_{ss}}{
k_{_B} T_{ss}} =
\frac{\left\langle {\bf p} \cdot c {\bf E} / m \right\rangle_{ss}}{
k_{_B} T_{ss}}
\label{KIKLG}
\end{equation}
where we have set ${\bf p}^2 / m = \langle {\bf p}^2 / m \rangle_{ss}
= (d - 1) k_{_B} T_{ss}$ and used that
$\langle {\bf p} \cdot {\bf F}^{\mbox{int}} \rangle_{ss} = 0$, similarly
to Eqs.(\ref{TIKss},\ref{phider}). We note that the right hand side of
Eq.(\ref{KIKLG}) equals the entropy production rate.

On the other hand,
\begin{equation}
\left\langle \kappa_{_{IE}} \right\rangle_{ss} =  (d - 1)
\left\langle \frac{{\bf p} \cdot c {\bf E} / m}{
{\bf p}^2 / m} \right\rangle_{ss} ~.
\label{KIELG}
\end{equation}
Since there is only one moving particle, one cannot
replace on the right hand side of (\ref{KIELG}) the average ratio
by the ratio of the averages. If one nevertheless does so and sets
$\langle {\bf p}^2 / m \rangle_{ss}
= d k_{_B} T_{ss}$, one obtains the expression
\begin{equation}
\left\langle \kappa_{_{IE}} \right\rangle_{ss} =  \frac{d - 1}{d} \frac{
\left\langle c {\bf p} / m \right\rangle_{ss} \cdot{\bf E}}{
k_{_B} T_{ss}} ~,
\label{KIEtemp}
\end{equation}
which differs from the expected value of the entropy production
rate of irreversible thermodynamics by an amount of
\begin{equation}
\frac{1}{d} \frac{
\left\langle c {\bf p} / m \right\rangle_{ss} \cdot{\bf E}}{
k_{_B} T_{ss}} ~,
\label{differ}
\end{equation}
which is of the same order as $\left\langle \kappa_{_{IE}} 
\right\rangle_{ss}$ itself.

The result of the above 
analysis is that for the IE Lorentz gas the average phase space
contraction and the entropy production rates are not equal, while 
they are in the case of the IK Lorentz gas.
At the same time, both the IE and the IK cases could be affected by the 
problem of unphysically long relaxation times.

Both difficulties can be overcome if many particles per unit volume 
are present in the system. Indeed, even non interacting particles would 
do, as the following three examples illustrate for the 
periodic (hard disk) Lorentz gas considered in Ref.\cite{MR}. 
In this case, there is an elementary cell (EC), whose copies tile the whole 
two dimensional plane, where each cell contains one fixed scatterer 
and one (or more) moving particles. Also, there are no external fields and 
thermostat couplings, so that this is an equilibrium system.  

{\bf (a)} In order to compute then the diffusion coefficient $D$ of
a point particle through the scatterers, we take a collection of $N$ 
particles (initially uniformly distributed
in each EC) $\{{\bf q}_i(0)\}_{i=1}^N$ and, similarly to what was
done in \cite{MR}, we let these particles evolve for a fixed time
$\zt$. Because of the periodic boundary conditions, the particle
density equals $N$ moving particles per EC at all times. 
Let us denote by ${\bf q}_i(\zt)$ the final 
position of the $i$-th particle trajectory and by ${\bf r}_i(\zt)$ 
the displacement ${\bf q}_i(\zt) - {\bf q}_i(0)$. The simulations
yield a value $D_{\mbox{sim}}$, which is related to $D$ by
\begin{equation}
D_{\mbox{sim}} \equiv \frac{\sum_{i=1}^N {\bf r}_i^2(\zt)}{N \zt} 
= D + \ze(N,\zt)
\label{diff}
\end{equation}
where $\ze(N,\zt)$ can be made arbitrarily small by taking
sufficiently large $N$ and $\zt$.\footnote{The best numerical
estimate for $D$ given in Ref.\cite{MR} is $D_{\mbox{num}}=0.2492(3)$.}
Since for the Lorentz gas the many independent initial conditions 
also represent the members of an ensemble, the extrapolation to
large $N$ can be seen here as one in ``ensemble size''. In other words,
taking the limit of large $N$ in Eq.(\ref{diff}), amounts to 
reproducing the microcanonical ensemble average 
$\langle {\bf r}^2(\zt) \rangle$ for any fixed $\zt > 0$.
How large does $\zt$ have to be for a given $N$, so that
$| \ze(N,\zt) |$ is smaller than a given $\varepsilon > 0$?
One finds that the larger $N$, the shorter $\zt$ can be
(see Figure 1 and Table 1). In particular, for $N=1$ no diffusive behaviour 
can be discerned at all in runs of $10^9$ collisions. For larger $N$, 
instead, we see that $D_{\mbox{sim}}$ values close to $D$ are obtained: the 
quicker the larger $N$ and the better the larger $\zt$.\footnote{This is 
qualitatively shown in Figure 1, where we have taken some special values of 
$N$. In Table 1 we also included intermediate values both for $N$ and time, 
so that larger oscillations about the general trend are present.} 
In the limit of very large $N$, diffusive behavior is then expected to 
become observable in times of about one collision time, i.e. the 
expected physical time.\footnote{Here and in the following we mainly
refer to the number of collisions per particle as a measure of the
elapsed time, because the convergence of the time averages of the phase
variables to the correct values is determined by this number. 
On the contrary, the actual values of the elapsed times have no special 
meaning by themselves since they depend on the scaling of the particle speed.} 

{\bf (b)} The same kind of behavior is observed in the same equilibrium
system for the internal potential part of the pressure defined by
\begin{equation}
p^\zF V = - \lim_{T\rightarrow\infty} \frac{1}{T} 
\sum_{\mbox{collisions}} {\bf p} \cdot \hat{\bf n} ~.
\label{potpre}
\end{equation}
Here the sum runs over all collisions between the $N$ moving particles and
the scatterers, which have taken place in the time $T$; {\bf p} is the
momentum of a colliding particle and $\hat{\bf n}$ is the outer 
normal to the scatterer at the collision point \cite{EM}.
Our results are given in Figure 2, while
the best numerical value for $p^\Phi V$ given in Ref.\cite{MR} is 
$0.5457(3)$. For a density of $N=1$ particle per EC
this value is obtained after $10^7$ collisions.

{\bf (c)}
The third example we want to mention is obtained from the previous 
Lorentz gas system, in the limit that the spacing between the scatterers
vanishes (the scatterers touch)
and that there is only one moving particle per EC. 
This way each moving particle is confined inside a triangular
like box, whose sides are arcs of three scatterer surfaces and we 
effectively simulate the dynamics of a {\it bona fide} 1-particle 
system in a box. This model was considered also in Ref.\cite{MR}
for the calculation of the pressure (see the smallest
value of $V$ in Figure 3 in Ref.\cite{MR}). 
The result is that a number of about $10^7$ collisions is 
necessary to obtain the value of the potential pressure
with the same accuracy of example (b), by following the motion of 
the confined particle. The diffusion coefficient is obviously zero 
and was not computed. 

The reason for a faster convergence of systems with larger $N$ is 
obviously that more collisions take place in the same (physical) time. 
Hence, we believe that also nonequilibrium systems will behave in this 
fashion. The effect of the external fields present in molecular 
dynamics simulations is discussed in point 1 of the next 
section in the context of the thermostatted Lorentz gas.

\section{Discussion}
{\bf 1.} Our results indicate that there is 
no conflict between dynamical systems theory
and Irreversible Thermodynamics. The entropy production rate of
Irreversible Thermodynamics is only obtained from the phase 
space contraction rate of a dissipative IE dynamical
system of $N$ particles if terms of order $O(1/N)$ can be neglected, 
i.e. in the thermodynamic limit. The difficulty arises 
not only for the IE case, but also for the IK case, although
a formal equality between the average phase space contraction 
and entropy production rates then holds for all $N$. Indeed, the time
needed for the corresponding averages to be compared with the 
${\bf J} \cdot {\bf X}$ of Irreversible Thermodynamics may be 
unphysically long if $N$ is too small. Unfortunately, the modern methods 
of nonequilibrium molecular dynamics (NEMD) are not too helpful
in this respect, not even for such a simple system as the
Lorentz gas, because the fields then used are well
beyond the linear regime.\footnote{This has been discussed, for
instance, in  pages 580-583 of Ref.\cite{CELS} in the context
of the van Kampen's objection to linear response theory \cite{VK}. 
Moreover, as pointed out also in Ref.\cite{CELS}, there may be regimes
in which {\it ``nonlinear corrections conspire to produce a linear effect
with modified slope''}. The authors of Ref.\cite{CELS}, then add
that this seems to be unlikely. However, it is consistent with
the results of Ref.\cite{LNRM}. For instance, Figure 2 of that paper,
shows a plateau in the conductivity as a function of field for large
fields (around $\approx 0.4$) which are already outside the linear regime.
Hence, the observation of linear relations
between thermodynamic forces and fluxes in NEMD does not necessarily
mean effective observation of the linear regime of Irreversible
Thermodynamics.}

The problem to correctly simulate systems close to equilibrium by NEMD
can be understood in terms of the model of Ref.\cite{LNRM}. In this  
model one moving particle, subject to a constant external field and
a Gaussian thermostat, experiences elastic collisions with
fixed circular scatterers, placed in a triangular lattice.
First, we observe that reducing the field makes it more and 
more difficult to obtain the transport coefficient. This is
illustrated in Figure 3 for the color conductivity $L$, defined
as the time average of the color current $J$ divided by the external field 
strength $F_e$. Secondly, reducing the field also introduces larger
errors in the calculation of the trajectory of the moving particle, so that 
a theoretical lower bound of about $10^{-6}$ for the field must be imposed.
In reality, a combination of the times needed for the runs and for
the precision that should be achieved, using double precision algorithms, 
imposes $10^{-4}$ as a lower bound for the field. At the same time, 
simple calculations show that a physical potential difference of $1 ~V/$cm 
corresponds in our model to an unobtainable field strength of at 
most $10^{-6}$, see also Ref.\cite{CELS,BGG}. Moreover, 
if one were to use the van Kampen argument \cite{VK} for the linear 
regime, then one should do simulations with $F_e \approx 10^{-35}$!

The question remains, in general, when dynamical systems results really 
correspond to those found for macroscopic systems. The possibly 
unphysically long times needed for small $N$ to define the observed
physical quantities as time 
averages makes then macroscopic concepts as temperature, pressure and 
their gradients (related to currents and thermodynamic forces),
which occur in Irreversible Thermodynamics, questionable.
This is even more relevant for thermostatted non-Hamiltonian systems,
such as the SLLOD equations, which describe viscous 
flows (see Ref.\cite{EM}).

\noindent
{\bf 2.} The necessity for the presence of many particles to obtain 
macroscopic behavior has been illustrated in terms of our results for
the periodic Lorentz gas. We observed that diffusion and potential 
pressure both reach their correct average values within short times if 
the system has a suffcient number of particles per EC. If, to the contrary, 
$N$ is small, large fluctuations have to be averaged out in the evolution 
of just the few particles present, which may take unphysically long times. 
From the mathematical point of view, instead, there is no problem 
because the time is then allowed to grow without bounds. In this sense,
there is a similarity between the problem we addressed here
and that posed by Poincar\'e's recurrence theorem and Boltzmann's 
H-theorem. 

We remark that it is not clear to us
what precise role chaoticity plays in obtaining macroscopic
behavior of the properties of the system, as defined by time averages.
While the presence
of many particles per unit volume is clearly necessary, the degree of
chaoticity needed to obtain physical behavior in physical times is unclear.
That is, it remains an open question to identify the precise conditions on 
the interactions among the particles in the system $-$which imply 
then a given 
degree of chaoticity$-$ so that physical behavior is obtained
in physical times. We note that for ensemble averages to describe the
properties of the system, a spreading of the ensemble in time is usually
necessary, which also requires a certain degree of chaoticity of
the system (see e.g. Ref.\cite{Gas1}).
Indeed, in the cases we investigated it is observed that the time 
average reaches a good value when also the ensemble of trajectories 
has covered enough of the phase space.

\noindent
{\bf 3.} Nevertheless, the theorems proved in Refs.\cite{RU,GR} for any
number of particles imply 
important new insights. In particular, in view of the fact that Ruelle 
\cite{RU} uses effectively the phase space contraction rate, his theorem 
on the positivity of the average phase space contraction rate is not 
restricted, due to its dynamical nature, to systems close to equilibrium.
Therefore, its applicability to macroscopic systems could
imply the positivity of the transport coefficients also far
from equilibrium, at least as long as there is strong chaoticity
and no vanishing Lyapunov exponents. This then could
lead to a proof of the positivity of the rheological viscosity
of a fluid, i.e., the shear rate dependent viscosity of a fluid under 
shear far from equilibrium, not just for the (hydrodynamic) Newtonian
viscosity of a fluid near equilibrium. Similarly, the cited proof of the 
Onsager reciprocal relations \cite{GR} 
seems really a proof of certain symmetry properties of the phase
space contraction rates of a given class of dynamical systems. 
Hence, they constitute an example of mathematical
results which, without physical motivation, would probably never 
have been considered in the context of dynamical systems. However, 
for large $N$, these mathematical results also have physical meaning.

\noindent
{\bf 4.} Our arguments on the relevance of time scales in the 
IE and IK models is further
strengthened by the recent results of Ref.\cite{CCP}, for the
Fermi-Pasta-Ulam problem. There, it is shown that
for sufficiently low energies there can be regions of
phase space with positive measure, in which the transient behavior
persists for arbitrarily large times and no equipartition of energy
between the degrees of freedom of the system is observed. However, 
lower and lower
energies are needed for these regions to exist as $N$ grows.
Thus, the growing of $N$ ensures that relaxation to the thermodynamic
behavior takes place within physical time scales.

\noindent
{\bf 5.} There is an additional reason why
the number of degrees of freedom plays a role in the IK models.
Consider one point particle in a one-dimensional space, with
singular potential barriers, e.g. let the potential energy be
$\Phi(x) = 1/|x|$. The equations of motion are then
\begin{equation}
\dot{x} = \frac{p}{m} ~; \quad \dot{p} = \frac{1}{x|x|} + F - \za p
\label{1dIK}
\end{equation}
where $\za p = F + 1/x|x|$ and $F$ is the force due to a constant
external field. But this means that
\begin{equation}
\dot{x} = \frac{p}{m} ~; \quad \dot{p} = 0
\label{1d0}
\end{equation}
which is trivially solved to give $x(t)=p(0)t/m + x(0)$. Therefore,
the particle can move {\em through} a singularity of the potential, if the 
signs of $x(0)$ and $p(0)$ are opposite. More generally, the IK models 
can in principle allow arbitrarily large values of the total energy, 
because the thermostat can do any required amount of work on the system, 
in order to keep the kinetic energy constant. This fact does not 
constitute a problem for the average quantities, if the isokinetic
distribution of Ref.\cite{EM} holds: in that case, the regions close to 
the singularities are given a weight which decreases exponentially
with the potential energy, under the assumption that terms of order 
$O(1/N)$ can be neglected (cf. Ref.\cite{EM}, Eqs.(5.25-5.28)).

\noindent
{\bf 6.} Our comments do not imply that the study of low
dimensional systems has no physical relevance from the point 
of view of transport theory.
There are many reasons why low dimensional dynamical systems
may correctly describe a given physical situation. For instance,
one could model by a dynamical system only the 
macroscopic behavior, like, e.g., in hydrodynamics. In that case only a 
few variables will play a significant role for most physical situations.
Also, recent studies show (see, e.g., Refs.\cite{Gas,Tel})
that the dynamics of chains of low-dimensional maps can be consistent
with macroscopic transport equations, if there are many copies of the 
same map in the chain, i.e. if the thermodynamic limit is taken 
in that sense. 

\noindent
{\bf 7.} Our arguments are relevant also for the problem of 
equivalence of ensembles in nonequilibrium statistical mechanics.
In particular, the idea that there may be several stationary
distributions which become equivalent in the thermodynamic limit, 
is here confirmed by the fact that both the IK and the IE dynamics give 
rise to stationary states with the same entropy production rates for 
$N \rightarrow \infty$. Similarly to Refs.\cite{Tel}, also here
the equivalence is based on the equality of entropy production rates, 
which determine the transport properties of the physical systems.
The idea of the equivalence of nonequilibrium ensembles is rather new 
and has been discussed earlier from a variety of point of
views in Refs.\cite{Tel,EQUI}.
 
\vskip 30pt

\noindent
{\Large \bf Acknowledgements} 
\vskip 10pt 
\noindent
This work has been supported by GNFM-CNR (Italy) and
through the EC contract ERBCHRXCT940460 (Stability and universality 
in classical mechanics).
EGDC gratefully acknowledges support from the US Department of Energy
under grant DE-FG02-88-ER13847.

\newpage
\centerline{\bf Figure captions}

\vskip 25pt
\noindent
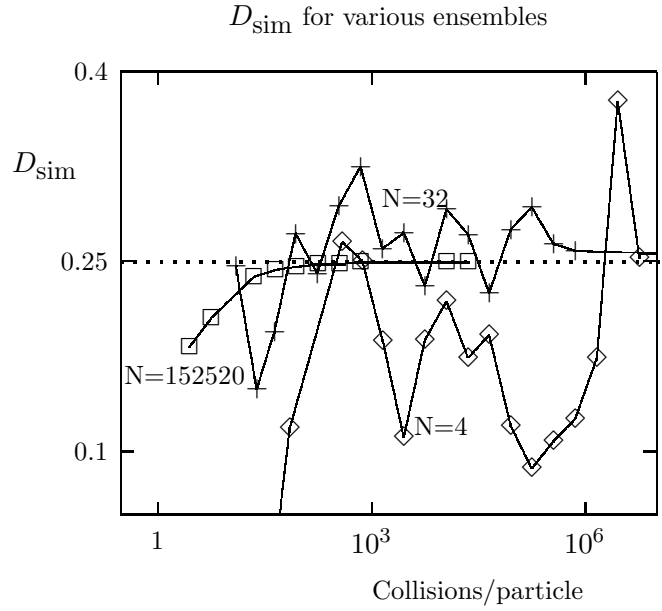
{\bf Figure 1.}
Diffusion coefficient $D_{\mbox{sim}}$ for the periodic (triangular) 
Lorentz gas with an interdisk surface to surface
spacing of $w=0.3$, and disk radius of $r=1$. The particle speed 
is also $1$. The various curves correspond to different ensembles made up
of $N$ initially uniformly distributed members. The horizontal axis 
represents the average number of collisions per particle, for each ensemble. 
For $N=4$ ($\Diamond$) we see
no convergence towards the value $D=0.2492$. For $N=32$ ($+$)
convergence appears after $10^7$ collisions. For $N=152520$ ($\Box$)
convergence takes place between $10$ and $100$ collisions. For 
$N \rightarrow \infty$, convergence is then expected to occur after about
$1$ collision.

\vskip 25pt
\noindent
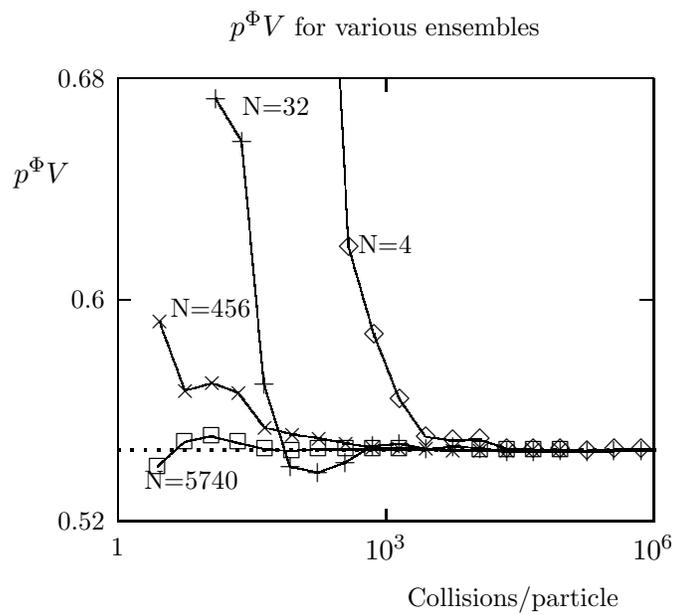
{\bf Figure 2.}
Potential pressure for the periodic (triangular) Lorentz gas
of Figure 1.
The various curves correspond to different ensembles made of $N$
members. The horizontal axis represents the average number of
collisions per particle for each ensemble. Clearly, all
curves converge to the numerically estimated 
value of $0.5457$ (cf. Ref.\cite{MR}) represented by the
dotted horizontal line, but the convergence is faster for larger $N$.

\vskip 25pt
\noindent
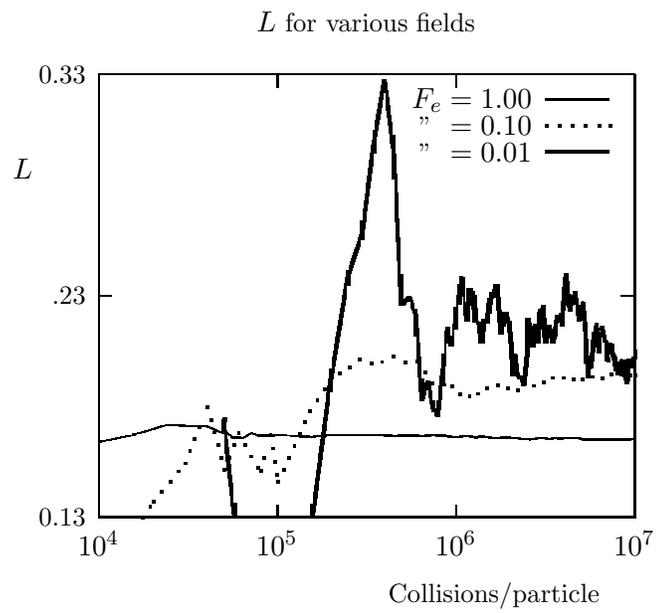
{\bf Figure 3.}
Here we see that the time it takes for the color conductivity $L$ in the
periodic, nonequilibrium Lorentz gas to reach its average value
increases as the field strength $F_e$ is decreased. For 
$F_e=1$, convergence to $0.165$ is obtained rather quickly, while for a 
field $F_e=0.01$ large fluctuations survive at relatively long times.
One should note that even our smallest field is much outside
the linear regime, which is below field strengths of $10^{-6}$.  

\vskip 50pt
\centerline{\bf Table captions}

\noindent
{\bf Table 1.}
Values of $\ze(N,\zt) = D_{\mbox{sim}} - D$, Eq.(\ref{diff}), for the
Lorentz gas of Figure 1 and for ensembles not
represented in that figure. The first column represents the time $\zt$
elapsed for each initial condition at the moment when Eq.(\ref{diff}) 
is evaluated, while the remaining columns refer to ensembles with $N$ 
initially uniformly distributed members. For $D$ we have used the best
numerical value $D_{\mbox{num}} = 0.2492$ given in Ref.\cite{MR}. 
The units are such that the radius of the scatterers, the particle speed
and mass are $1$, while the distance between the centers of two nearest 
neighbor scatterers is $2.3$.  
This table can also be used to choose $N$ and $\zt$ 
to get a desired accuracy. The actual values of the time are not
important: only the relative times are, because the time units depend
on the velocity scaling.

\newpage
\begin{table}
\begin{center}
Table 1
\vskip 3cm

\vskip 5cm
Figure 3
\end{center}
\end{figure}


\newpage   

\begin{thebibliography}{99}
\addcontentsline{toc}{chapter}{Bibliography}
\bibitem{RU} D. Ruelle, 1996, {\it Positivity of entropy production
in nonequilibrium statistical mechanics}, J. Stat. Phys., {\bf 85}, 1
\bibitem{RMlarge} L. Rondoni and G. P. Morriss, 1997, {\it
Applications of periodic orbit theory to N-particle systems}, J. Stat.
Phys., {\bf 86}, 991
\bibitem{RC} L. Rondoni and E.G.D. Cohen, 1997, {\it Orbital
measures in nonequilibrium statistical mechanics: the Onsager relations},
archived in http://xxx.lanl.gov/chao-dyn \#9708019
\bibitem{GR} G. Gallavotti and D. Ruelle,
1996, {\it SRB states and nonequilibrium statistical mechanics}, 
archived in mp\_arc @ math.utexas.edu, \#96-645
\bibitem{GG1} G. Gallavotti, {\it Chaotic hypothesis:
Onsager reciprocity and fluctuation dissipation theorem}, J.
Stat. Phys., {\bf 84}, 899 (1996)
\bibitem{GG2} G. Gallavotti, {\it Extension of
Onsager's reciprocity to large fields and the chaotic hypothesis}, 
Phys. Rev. Lett., {\bf 78}, 4334 (1996)
\bibitem{SA} D. Szasz, {\it Boltzmann's ergodic hypothesis, a conjecture
for centuries?}, Studia Scie. Math. Hungarica, {\bf 31}, 299 (1996)
\bibitem{GA} K.F. Gauss, {\it \"Uber ein neues allgemeines Grundgesetz
der Mechanik}, J. Reine Angew. Math., {\bf IV}, 232 (1829)
\bibitem{EM} D.J. Evans and G.P. Morriss, {\it Statistical mechanics of
nonequilibrium liquids}, Academic Press, London (1990)
\bibitem{BEC} A. Baranyai, D.J. Evans and E.G.D. Cohen, 1993,
{\it Field-dependent conductivity and diffusion in a
two-dimensional Lorentz gas}, J. Stat. Phys., {\bf 70}, 1085
\bibitem{MR} G. P. Morriss and L. Rondoni, 1994, {\it Periodic orbit 
expansions for the Lorentz gas}, J. Stat. Phys., {\bf 75}, 553
\bibitem{CELS} N. I. Chernov, G. L. Eyink, J.L. Lebowitz and Ya. Sinai,
1993, {\it Steady state electric conductivity in the periodic Lorentz gas},
Comm. Math. Phys., {\bf 154}, 569
\bibitem{VK} N.G. van Kampen, {\it The case against linear response
theory}, Physica Norvegica, {\bf 5}, 279 (1971)
\bibitem{LNRM} J. Lloyd, M. Niemeyer, L. Rondoni, and G.P. Morriss, 
{\it The nonequilibrium Lorentz gas}, CHAOS, {\bf 5}, 536 (1995)
\bibitem{BGG} F. Bonetto, G. Gallavotti and P.L. Garrido, {\it Chaotic 
principle: an experimental test}, Physica D,
{\bf 105} 226 (1997)
\bibitem{Gas1} P. Gaspard, {\it Hydrodynamic modes as singular eigenstates
of the Liouvillian dynamics: Deterministic diffusion}, Phys. Rev. E 
{\bf 53}, 4399 (1996)
\bibitem{CCP} L. Casetti, M. Cerruti-Sola, M. Pettini and E.G.D. Cohen, 
{\it The Fermi-Pasta-Ulam problem revisited}, Phys. Rev. E, {\bf 55},
6566 (1997)
\bibitem{Gas} S. Tasaki and P. Gaspard, {\it }, J. Stat. Phys., {\bf 81},
935 (1995)
\bibitem{Tel} T. Tel, J. Vollmer and W. Breymann, {\it Transient Chaos:
the origin of transport in driven systems}, Europhys. Lett., {\bf 35}(9),
659 (1996). W. Breymann, T. Tel, and J. Vollmer, {\it
Entropy Production in Open Dynamics Systems},
Phys. Rev. Lett., {\bf 77}, 2945 (1996)
\bibitem{EQUI} D.J. Evans and S. Sarman, {\it Equivalence of thermostatted
nonlinear responses}, Phys. Rev. E, {\bf 48}, 65 (1993).
D.J. Evans, {\it The equivalence of Norton and Thevenin ensembles},
Mol. Phys., {\bf 80}, 221 (1993).
G. Gallavotti, {\it Ergodicity, ensembles, irreversibility
in Boltzmann and beyond}, J. Stat. Phys. {\bf 78}, 1571 (1995).
G. Gallavotti and E.G.D. Cohen, {\it Dynamical ensembles
in stationary states}, J. Stat. Phys., {\bf 80}(5/6), 931 (1995).
L. Rondoni and G.P. Morriss, {\it Equivalence of
``nonequilibrium'' ensembles for simple maps}, Physica A, {\bf 233},
767 (1996). G. Gallavotti, {\it Dynamical ensemble equivalence in
fluid mechanics}, archived in chao-dyn/9605006. G. Gallavotti,
{\it New methods in nonequilibrium gases and fluids}, archived in
mp\_arc 96-533.
\end{thebibliography}
\end{document}